\documentclass[useAMS,usenatbib]{mn2e}
\usepackage{epsfig}
\usepackage{amsmath}
\usepackage{graphicx}
\usepackage[colorlinks, citecolor = blue]{hyperref}
\usepackage{times}
\usepackage{abbreviations}
\usepackage{breakcites}
\usepackage{subfig}

\newcommand{\HII}{\mbox{H II}}
\newcommand{\OIII}{\mbox{[O III]}}
\newcommand{\OII}{\mbox{[O II]}}

\newcommand{\NII}{\mbox{[N II]}}
\newcommand{\SII}{\mbox{[S II]}}
\newcommand{\Ha}{H$\alpha$}
\newcommand{\Hb}{H$\beta$}

\newcommand{\NIIHa}{\NII/H$\alpha$}
\newcommand{\SIIHa}{\SII/H$\alpha$}

\newcommand{\OIIIHb}{\OIII/H$\beta$}

\title[Star Formation, Shocks and AGN in NGC~613]{Dissecting Galaxies: Separating Star Formation, Shock Excitation and AGN Activity in the Central Region of NGC~613}

\author[R. L. Davies et al.]{Rebecca L. Davies$^{1,2}$\thanks{Email: \href{mailto:rdavies@mpe.mpg.de}{rdavies@mpe.mpg.de}}, Brent Groves$^{1}$, Lisa J. Kewley$^{1}$, Anne M. Medling$^{1,3,4}$, \newauthor Prajval Shastri$^5$, Jaya Maithil$^6$, Preeti Kharb$^7$, Julie Banfield$^{1,8}$, Fergus Longbottom$^{1}$, \newauthor Michael A. Dopita$^{1}$, Elise J. Hampton$^1$, Julia Scharw\"achter$^9$, Ralph Sutherland$^1$, \newauthor Chichuan Jin$^2$, Ingyin Zaw$^{10}$, Bethan James$^{11}$, St\'ephanie Juneau$^{12}$ \\ \\
$^1$Research School of Astronomy and Astrophysics, Australian National University, Canberra, ACT 2611, Australia \\
$^2$Max-Planck-Institut f\"{u}r Extraterrestrische Physik, Giessenbachstrasse, D-85748 Garching, Germany \\
$^3$Cahill Center for Astronomy and Astrophysics, California Institute of Technology, MS 249-17, Pasadena, CA 91125, USA \\
$^4$Hubble Fellow \\
$^5$Indian Institute of Astrophysics, Sarjapur Road, Bengaluru 560034, India \\
$^6$Department of Physics \& Astronomy, University of Wyoming, WY 82071, USA \\
$^7$National Centre for Radio Astrophysics - Tata Institute of Fundamental Research, Pune University Campus, Post Bag 3, Ganeshkhind Pune 411007, India \\
$^8$ARC Centre of Excellence for All-Sky Astrophysics (CAASTRO) \\
$^9$Gemini Observatory, Northern Operation Center, 670 N. A'ohoku Place, Hilo, HI 96720, USA \\
$^{10}$New York University Abu Dhabi, P.O. Box 129188, Abu Dhabi UAE \\
$^{11}$Institute of Astronomy, Cambridge University, Madingley Road, Cambridge CB3 0HA, UK \\
$^{12}$CEA-Saclay, DSM/IRFU/SAp, 91191 Gif-sur-Yvette, France} 

\begin{document}

\maketitle

\begin{abstract}
The most rapidly evolving regions of galaxies often display complex optical spectra with emission lines excited by massive stars, shocks and accretion onto supermassive black holes. Standard calibrations (such as for the star formation rate) cannot be applied to such mixed spectra. In this paper we isolate the contributions of star formation, shock excitation and active galactic nucleus (AGN) activity to the emission line luminosities of individual spatially resolved regions across the central \mbox{3~$\times$~3 kpc$^2$} region of the active barred spiral galaxy NGC~613. The star formation rate and AGN luminosity calculated from the decomposed emission line maps are in close agreement with independent estimates from data at other wavelengths. The star formation component traces the B-band stellar continuum emission, and the AGN component forms an ionization cone which is aligned with the nuclear radio jet. The optical line emission associated with shock excitation is cospatial with strong $H_2$ and \mbox{[Fe II]} emission and with regions of high ionized gas velocity dispersion \mbox{($\sigma \ga 100$ km s$^{-1}$)}. The shock component also traces the outer boundary of the AGN ionization cone and may therefore be produced by outflowing material interacting with the surrounding interstellar medium. Our decomposition method makes it possible to determine the properties of star formation, shock excitation and AGN activity from optical spectra, without contamination from other ionization mechanisms. 
\end{abstract}


\section{Introduction}
The combination of spectral diagnostics and spatial information makes integral field spectroscopy a powerful tool for unveiling the physical processes driving galaxy evolution. Over the past 40 years, astronomers have used the luminosities and ratios of emission lines as diagnostics for a range of galaxy properties, including the principle power source(s) \citep{Baldwin81, Veilleux87, Ke01a, Ke06}, star formation rate (SFR; e.g. \citealt{Kennicutt98, Jansen01, Brinchmann04, Ke04, Moustakas06, Kennicutt12}), AGN bolometric luminosity \citep[e.g.][]{Heckman04, Lamastra09}, and the metallicity \citep[e.g.][]{Pagel79, McGaugh91, Zaritsky94, Pilyugin01, KD02, Kobulnicky04, Pilyugin05, Liang07, Yin07, Ke08}, ionization parameter \citep[e.g.][]{KD02, Levesque14}, temperature and density of the ionized gas \citep{Osterbrock06}. These diagnostics have been extensively applied to data from large single fibre galaxy surveys, revealing important relationships between stellar mass and the global SFR (SFR main sequence; \citealt{Daddi07, Elbaz07, Noeske07}), the metal content (mass-metallicity relation; \citealt{Tremonti04}) and the fraction of galaxies hosting an AGN \citep{Ka03, Baldry04}. More recently, emission line diagnostics have also been applied to integral field observations of galaxies, providing insights into how the metal content \citep[e.g.][]{Rich12, Sanchez14, Ho15}, gas properties, and principle ionization mechanism(s) \citep[e.g.][]{Rich11, Davies14b, Davies14a, Dopita14, Ho14, Leslie14, Belfiore16, Davies16decomp} vary within individual galaxies. The addition of spatial information has opened up many new avenues in the study of galaxy evolution - for example, making it possible to place strong constraints on the inflow and outflow histories of galaxies \citep[e.g.][]{Ho15} and allowing us to directly observe the suppression and triggering of star formation by AGN feedback \citep[e.g.][]{Croft06, Elbaz09, CanoDiaz12, Rauch13, Cresci15, Salome15}. 

Emission line diagnostics are powerful probes of the physical conditions and processes occurring within galaxies, but many diagnostics require stringent assumptions about the nature of the ionization mechanisms contributing to the observed spectra. For example, the \mbox{\Ha\ $\lambda$6563} luminosity scales with the number of ionizing photons produced in massive stars and is therefore commonly used as an SFR indicator. However, \Ha\ can also be collisionally excited and is therefore not an accurate SFR indicator in the presence of shocks or the ionizing radiation field of an AGN. Similarly, many of the emission line ratios used in metallicity and ionization parameter diagnostics depend on the shape of the ionizing radiation field and therefore these diagnostics are calibrated for spectra dominated by a single ionization mechanism (usually star formation; e.g. \citealt{KD02}). Galaxies undergoing rapid phases of evolution (through processes such as galaxy-galaxy interactions and outflows) are prime laboratories for studying galaxy evolution, but often have multiple ionization mechanisms contributing to their optical line emission \citep[e.g.][]{Rich11, Leslie14, Rich15}. It has therefore been very difficult to calculate the star formation rates and metallicities of these galaxies in the past. Separating the line emission of these galaxies into contributions from individual ionization mechanisms would make it possible to calculate the star formation rates and metallicities without contamination from other ionization mechanisms.

The \NIIHa\ vs. \OIIIHb\ diagnostic diagram is commonly used to separate galaxies dominated by star formation from galaxies dominated by more energetic ionization mechanisms\footnote{A `more energetic' ionization mechanism is one for which the typical photon energy (for photoionization) or electron energy (for collisional ionization) is greater than that produced by young stars. This in general includes the presence of photons with energies $\ga$42 eV, not produced in massive main-sequence stars.} \citep{Baldwin81, Veilleux87, Ke01a, Ka03}. Galaxies dominated by star formation lie along a chemical abundance sequence \citep{Dopita86, Dopita00}. Galaxies with significant contributions from ionization mechanisms more energetic than star formation lie along the AGN branch of the diagnostic diagram, which spans from the enriched end of the star-forming sequence towards larger \NIIHa\ and \OIIIHb\ ratios. (The label `AGN branch' is somewhat of a misnomer, because shock excitation and evolved stellar populations can also produce line ratios along this branch of the diagram; e.g. \citealt{Fosbury78, Binette94, Stasinska08, Rich11}.) The presence of a more energetic ionization mechanism increases the collisional excitation rate in the nebula, enhancing the ratios of the forbidden \NII\ and \OIII\ lines to the \Ha\ and \Hb\ recombination lines. 

To first order, the greater the fraction of the line emission associated with shock excitation and/or AGN activity, the greater the enhancement in the diagnostic line ratios and the further along the AGN branch a galaxy will lie. The AGN branch is therefore often referred to as a global mixing sequence. \citet{Ke06} established the `star-forming distance' ($d_{SF}$, the distance of a galaxy spectrum from the star-forming sequence of the \NIIHa\ vs. \OIIIHb\ diagnostic diagram) as a metric for the positions of galaxies along the global mixing sequence. Subsequently, \citet{Kauffmann09} used the positions of single fibre spectra of galaxies along the global mixing sequence to estimate the fractional contribution of star formation to their \OIII\ luminosities.

With the advent of integral field spectroscopy, it is now possible to conduct detailed studies of how the diagnostic line ratios and the relative contributions of different ionization mechanisms vary within individual galaxies. Integral field studies of AGN host galaxies have revealed tight line ratio sequences between the star formation and AGN dominated regions of the \NIIHa\ vs. \OIIIHb\ diagnostic diagram (\citealt{Scharwaechter11, Davies14b, Davies14a, Dopita14, Belfiore15, Davies16decomp}). The line ratios often vary radially, from AGN-like line ratios in the centres of galaxies to \HII-like line ratios at larger galactocentric distances. These radial trends suggest that the line ratio variations are primarily driven by variations in the relative contributions of star formation and AGN activity across AGN host galaxies.

Similarly, tight mixing sequences between star formation and shock excitation have been observed in galaxies impacted by processes such as galaxy-galaxy interactions and outflows \citep[e.g.][]{Rich11, Rich15}. Shock excitation and AGN activity are not well separated on the \NIIHa\ vs. \OIIIHb\ diagnostic diagram but do occupy distinctly different regions on the \SIIHa\ vs. \OIIIHb\ diagnostic diagram \citep{Veilleux87, Ke06}. Seyfert AGN are characterised by larger \OIIIHb\ ratios and smaller \SIIHa\ ratios than shocks. In galaxies with mixing sequences between star formation and shock excitation, the spaxels with the smallest line ratios are often found in the nuclear regions where the star formation is strongest, and the line ratios increase towards larger galactocentric distances as the relative contribution of star formation (shock excitation) to the line emission decreases (increases).

Some galaxies display line ratio distributions which are not consistent with starburst-AGN or starburst-shock mixing sequences, but are produced by significant contributions from star formation, shock excitation \textit{and} AGN activity \citep[e.g.][]{Leslie14, Rich15}. When three ionization mechanisms are present, there is no clear correspondence between the diagnostic line ratios and the relative contributions of different ionization mechanisms to the observed line emission. In these cases, an alternate method must be used to separate the spectra into their underlying components.

\citet{Davies16decomp} presented a new method for separating the spectra of individual spaxels of integral field datacubes into contributions from different ionization mechanisms. We demonstrated that the line luminosities of the spectra of individual spaxels along the starburst-AGN mixing sequences of NGC~5728 and NGC~7679 are well represented by linear superpositions of the line luminosities of representative \HII\ region and AGN narrow line region (NLR) basis spectra. We separated the \Ha, \Hb, \NII, \SII, \OII\ and \OIII\ luminosities of each spaxel into contributions from star formation and AGN activity, and presented maps of the \Ha\ and \OIII\ emission associated with each of the ionization mechanisms. 

Here, we extend this method to galaxies with significant contributions from three ionization mechanisms. Specifically, we test how well our method can separate emission associated with star formation, shock excitation, and AGN activity in the local AGN host galaxy NGC~613, which shows clear evidence for all three of these ionization mechanisms. Section \ref{sec:obs} of this paper describes our observations and data processing, and discusses the ionization mechanisms contributing to the emission line signature of NGC~613. Our method is outlined in Section \ref{sec:decomp}, and maps of the emission associated with each of the ionization mechanisms are presented in Section \ref{sec:results}. The caveats associated with our method are discussed in Section \ref{sec:discussion} and our conclusions are summarised in Section \ref{sec:conc}.

\section{NGC~613}
\label{sec:obs}

\subsection{Observations}
This paper is based on integral field data for NGC~613, a nearby ($D$~=~26.4 Mpc, assuming \mbox{$H_0$ = 73.0 km s$^{-1}$ Mpc$^{-1}$}; \citealt{Tully09, Nasonova11}) SBbc galaxy \citep{deVaucouleurs91} with clear evidence for current star formation, shock excitation \citep{FalconBarroso14}, and AGN activity \citep{Goulding09}. The kiloparsec-scale radio structure of the galaxy (as imaged by the NRAO VLA Sky Survey at 1.4~GHZ) reveals enhanced star formation activity at the ends of the bar, as well as enhanced nuclear emission. 

NGC~613 was observed as part of the Siding Spring Southern Seyfert Spectroscopic Snapshot Survey (S7), an integral field survey of 130 nearby (z~$\la$~0.02), southern (declination \mbox{$<$ 10$^\circ$}) AGN host galaxies conducted using the ANU 2.3m telescope at Siding Spring Observatory \citep{Dopita15}. We observed the central \mbox{38~$\times$~25 arcsec$^2$} region of NGC~613 at a position angle of 130$^\circ$ for a total of 2400s. The observations were conducted using the high resolution red grating (covering $\lambda$~=~530-710 nm, with a spectral resolution of R = 7000 at $\lambda$ = 620 nm, corresponding to a velocity resolution of 43 km s$^{-1}$) and a lower resolution blue grating (covering $\lambda$~=~340-570 nm, with a spectral resolution of R = 3000 at \mbox{$\lambda$ = 468 nm}, corresponding to a velocity resolution of \mbox{100 km s$^{-1}$}). Full details of the observations can be found in \citet{Dopita15}.

\begin{figure}
\centerline{\includegraphics[scale = 0.45, clip = true, trim =  20 0 20 0]{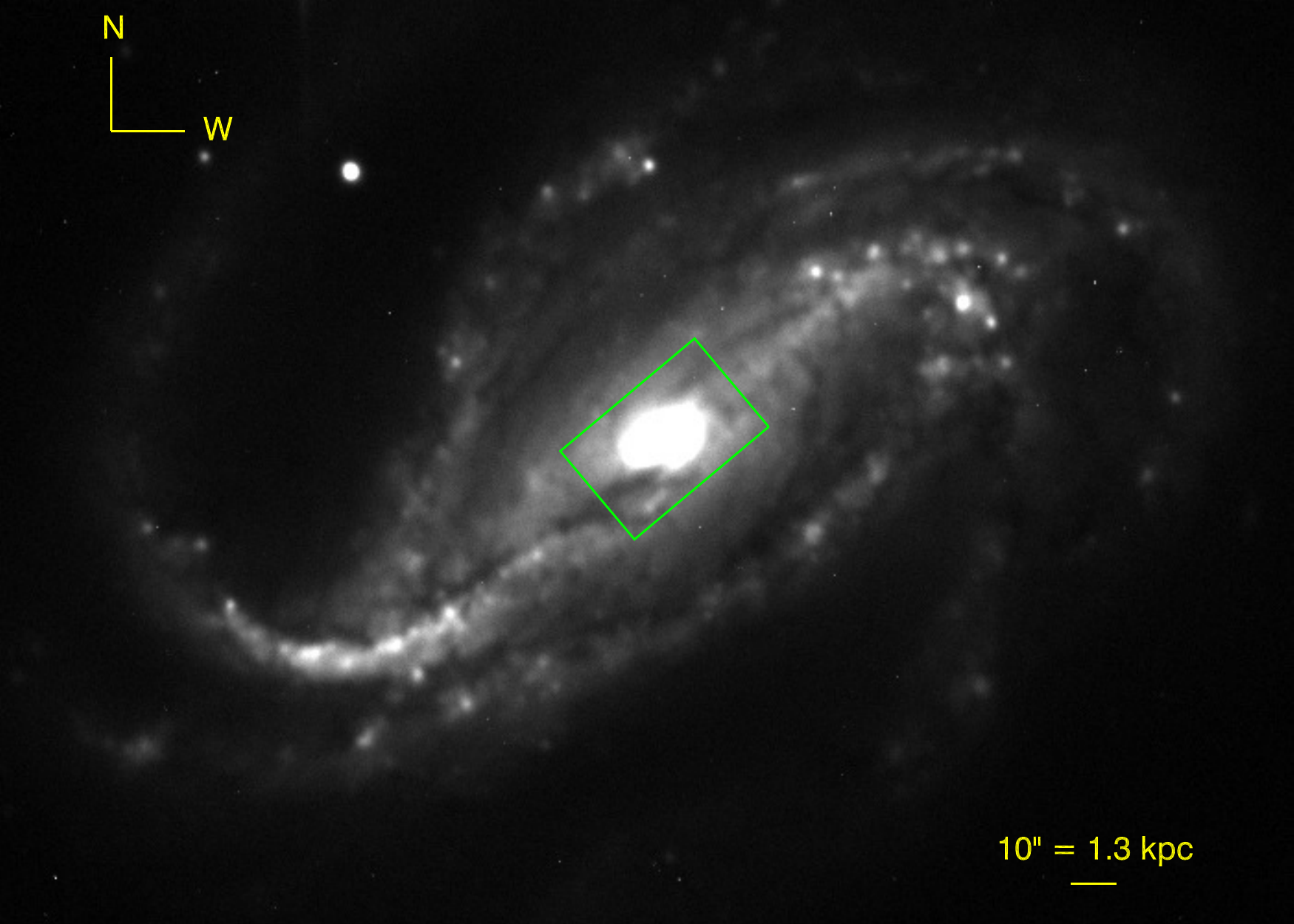}}
\caption{CTIO 0.9m B-band image of NGC~613 \citep{Eskridge02} with the 38~$\times$~25 arcsec$^2$ WiFeS footprint overlaid in green. The yellow scale bar has a length of 10 arcsec, corresponding to 1.3~kpc at a distance of 26.4 Mpc.}
\label{fig:dss}
\end{figure}

Figure \ref{fig:dss} shows the WiFeS footprint overlaid on a Cerro Tololo Inter-American Observatory (CTIO) 0.9m B-band image of NGC~613 \citep{Eskridge02}. The angular effective radius of NGC~613 is 65~arcsec, and therefore our S7 observations cover only the central region of the galaxy. The Full Width at Half Maximum (FWHM) of the spatial Point Spread Function (PSF) for the WiFeS observations was 1.5 arcsec, corresponding to a physical resolution of $\sim$190~pc. The AGN NLR is well resolved in our observations (see Figure \ref{fig:line_ratio_maps}).

\subsection{Data Reduction and Processing}
\label{subsec:reduction}

The data were reduced using the Python pipeline \textsc{PyWiFeS}, which performs overscan and bias subtraction, interpolates over bad CCD columns, removes cosmic rays, derives a wavelength solution from the arc lamp observations, performs flat-fielding to account for pixel-to-pixel sensitivity variations across the CCDs, re-samples the data and errors into cubes and finally performs telluric correction and flux calibration. Absolute photometric calibration of the data cubes was performed using the Space Telescope Imaging Spectrograph (STIS) spectrophotometric standard stars\footnote{Fluxes available at: \newline {\url{www.mso.anu.edu.au/~bessell/FTP/Bohlin2013/GO12813.html}}}. The wavelength solutions have a root mean square (RMS) error of \mbox{$\la$ 0.05 \AA} across the entire detector \citep{Childress14}, and the typical absolute spectrophotometric accuracy of the flux calibration is 4 per cent.  

Emission line fitting was performed using the \textsc{IDL} toolkit \textsc{LZIFU} \citep{Ho16}. \textsc{LZIFU} feeds each spectrum into the penalized pixel fitting routine (\textsc{pPXF}; \citealt{Cappellari04}), which fits the stellar continuum emission as a linear combination of the model stellar spectra from \citet{GonzalezDelgado05}. The errors output by \textsc{PyWiFeS} are taken into account when calculating the penalized likelihood of each template combination. 

\begin{figure*}
\captionsetup[subfigure]{labelformat=empty}
\subfloat[]{\includegraphics[scale = 1]{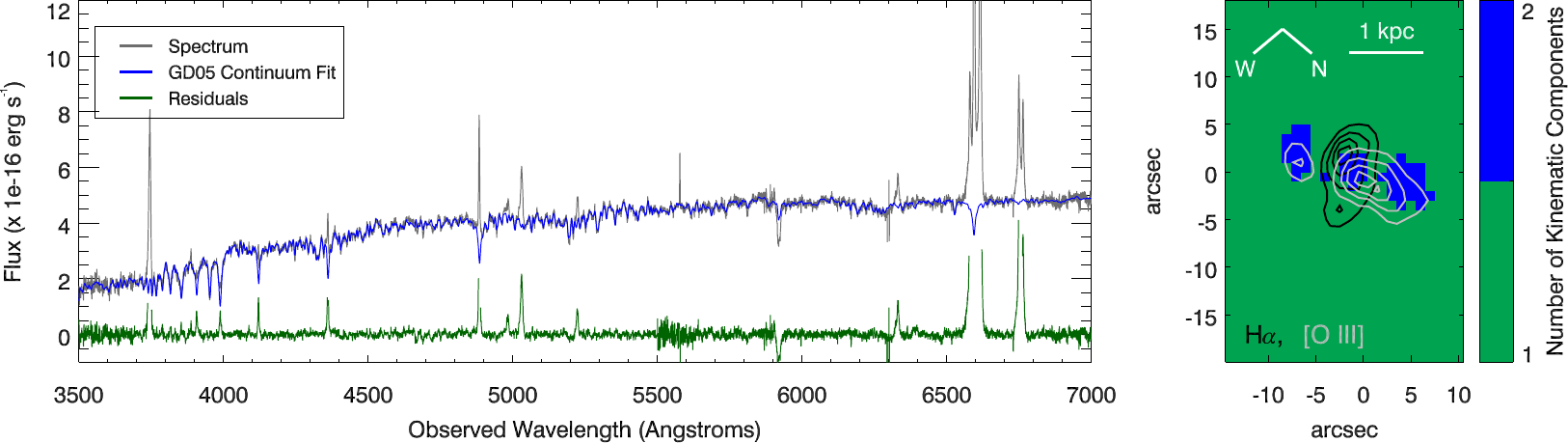}} \\
\subfloat[]{\includegraphics[scale = 1, clip = true, trim = 0 83 0 90]{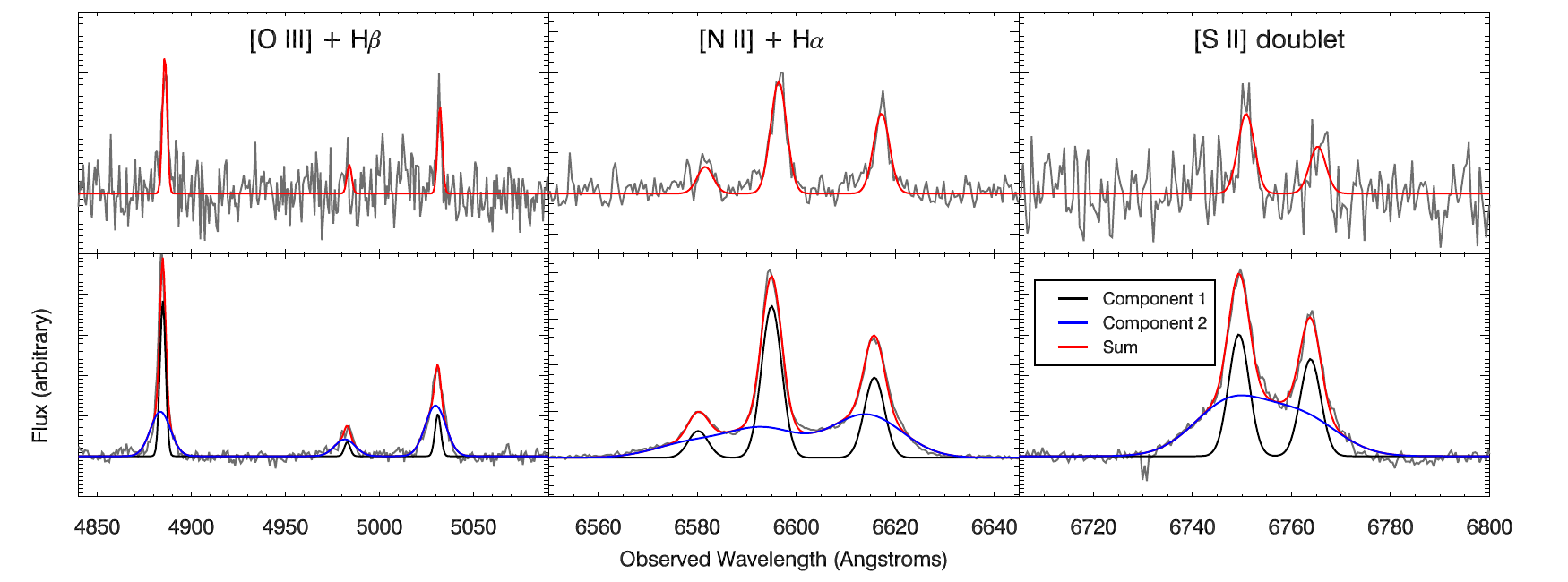}}
\caption{Top left: \citet{GonzalezDelgado05} stellar continuum fit (blue) to the nuclear spectrum of NGC~613 (grey). The residuals of the fit are shown below in green (regions around the centres of the strongest emission lines have been masked). Top right: Map showing the number of kinematic components used in this analysis, for each spaxel in our WiFeS observation of NGC~613. The black and grey contours trace the ridge lines for emission at levels of 30, 50, 70 and 90 per cent of the peak \Ha\ and \OIII\ emission, respectively. Bottom: Fits to the continuum-subtracted \OIII+\Hb, \NII+\Ha\ and \SII\ complexes for example spaxels requiring one (top row) and two (bottom row) kinematic components.}
\label{fig:fitting_demo}
\end{figure*} 

Our primary aim in fitting the continuum is to correct for the underlying stellar continuum and its features, in particular the Balmer absorption lines which impact the determined emission line strengths. The top left panel of Figure \ref{fig:fitting_demo} shows the nuclear spectrum of NGC~613 in grey, the best fit superposition of stellar templates in blue and the fitting residuals in green (regions around the centres of the strongest emission lines have been masked). The stellar population fit reproduces the strong Balmer absorption series blueward of 4000\AA, indicating that the fit provides a reasonable estimate of the stellar absorption under the \Ha\ and \Hb\ emission lines.

The stellar continuum fit is subtracted from the original spectrum, leaving the emission line spectrum. The emission line fluxes and ionized gas kinematics are extracted from the emission line spectra using the IDL Levenberg-Marquardt least squares fitting routine \textsc{MPFIT} \citep{Markwardt09}. We do not detect any broad \mbox{($\sigma \ga$1000 km s$^{-1}$)} \Ha\ or \Hb\ emission and therefore we assume that the emission lines are produced in \HII\ regions, shock excited regions and/or the AGN NLR. Each emission line spectrum is fit three times independently - once with a single component, once with two components and once with three components. Each component consists of a set of Gaussian functions (one associated with each emission line) which all have the same width (velocity dispersion) and velocity offset (relative to the input spectroscopic redshift).  The \textsc{PyWiFeS} errors are propagated through the emission line fitting process to produce the error on the flux measured for each emission line in each kinematic component of each spaxel. 

\begin{figure*}
\includegraphics[scale = 1.15, clip = true, trim =  0 70 190 20]{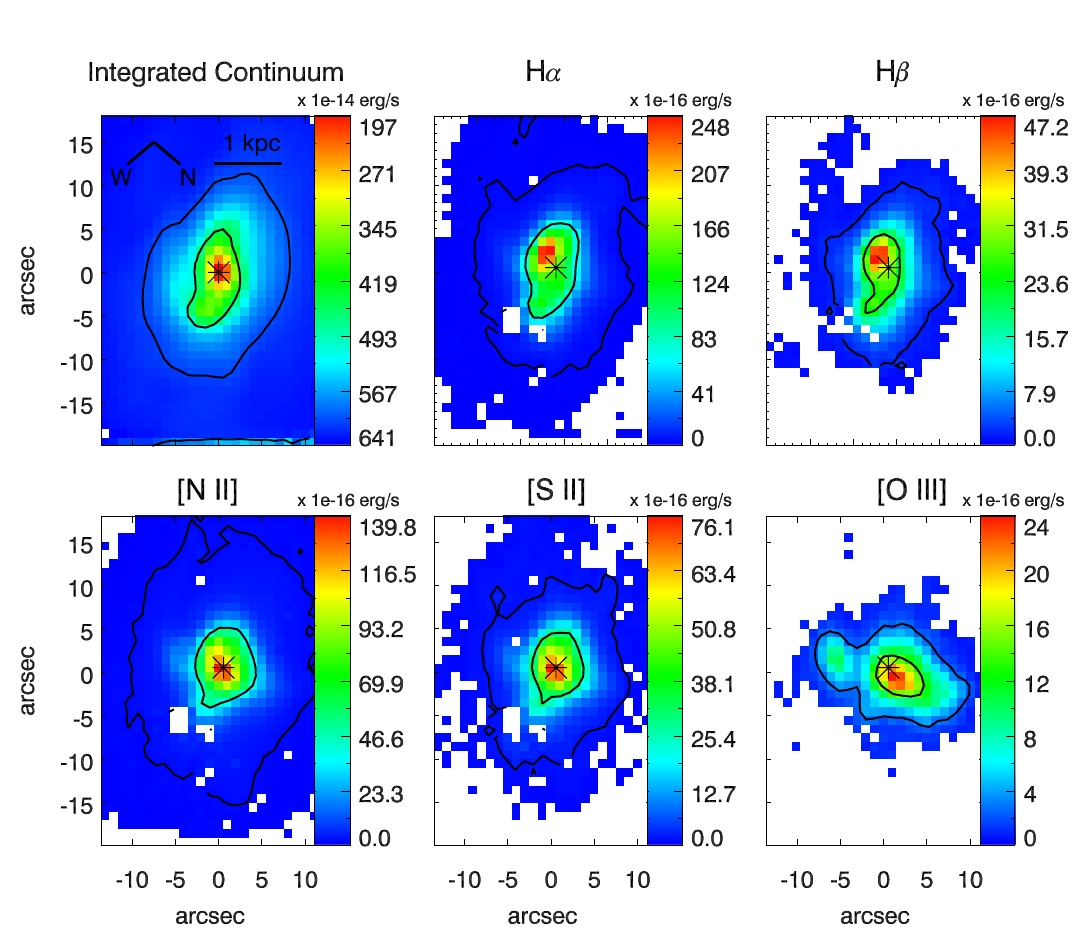}
\caption{Maps of the integrated continuum, \Ha, \Hb, \NII, \SII\ and \OIII\ emission across the observed region of NGC~613. The emission line flux maps show the total flux (summed across all kinematic components) in the relevant emission line, for each spaxel in which the relevant line is detected with \mbox{S/N $>$ 5}. Black asterisks on the images mark the peak of the integrated continuum emission (assumed to be the galaxy centre). The black contours indicate the regions containing 68 per cent (1$\sigma$) and 95 per cent (2$\sigma$) of the flux in each image.}
\label{fig:flux_maps}
\end{figure*}

An artificial neural network (ANN) called \textsc{LZComp} (Hampton et al. 2017, accepted) was employed to determine which of the emission line fits (with 1, 2 or 3 kinematic components) best represents the emission line spectrum of each spaxel. The ANN uses the signal-to-noise ratios (S/N) and reduced $\chi^2$ values of the emission line fits to determine how many kinematic components can be robustly detected in a given spectrum. Weak kinematic components (which account for a small percentage of the total emission line flux) can be recovered even if they are only robustly detected in one or two of the strongest emission lines. However, for the majority of this study we are only able to use components for which all the required diagnostic lines (\Ha~$\lambda$6563, \Hb~$\lambda$4861, \NII~$\lambda$6583, \mbox{\SII~$\lambda \lambda$6716, 6731} and \OIII~$\lambda$5007) are detected with \mbox{S/N $>$ 5}. We take each spaxel assigned more than one kinematic component and check whether all five of the diagnostic lines in all of the kinematic components are detected with \mbox{S/N $>$ 5}. If not all of the diagnostic lines pass the S/N threshold, the number of kinematic components assigned to that spaxel is reduced by one. This maximises the number of components we are able to use in our study whilst having minimal impact on the total emission line fluxes measured for each spaxel.

The top right panel of Figure \ref{fig:fitting_demo} shows a map of the number of kinematic components assigned to each spaxel for this study. Approximately five per cent of the spaxels are assigned two components, with all remaining spaxels assigned one component. None of the spaxels have all the required emission lines detected in three kinematic components.

The bottom panel of Figure \ref{fig:fitting_demo} shows the fits to the \OIII+\Hb, \NII+\Ha\ and \SII\ complexes for a spaxel assigned one kinematic component (top row) and a spaxel assigned two kinematic components (bottom row). The spaxel assigned one kinematic component displays fairly regular, narrow line profiles. The \NII\ and \Ha\ line profiles show evidence for a second, broader,  blue-shifted component but it is relatively weak and is not seen in the \SII, \OIII\ or \Hb\ lines. On the other hand, the spaxel assigned two kinematic components shows very irregular and asymmetric line profiles, with clear evidence for a prominent broad component which is blue-shifted with respect to the narrow component.

\subsection{Evidence for Three Ionization Mechanisms in the S7 Spectra of NGC~613}
\label{subsec:ionization_mechanisms}
NGC~613 is an excellent test case for our method because it shows clear evidence for star formation, shock excitation and AGN activity \citep[see e.g.][]{Hummel87, Hummel92, Goulding09, FalconBarroso14}. In the following sections we analyse continuum and line emission maps, line ratio maps, diagnostic diagrams and kinematic information, to demonstrate that all three ionization mechanisms contribute significantly to the line emission within the observed region of NGC~613.

\subsubsection{Continuum and Line Emission Maps} 
\label{subsubsec:flux_maps}
Figure \ref{fig:flux_maps} shows maps of the integrated continuum flux and the flux in the five diagnostic emission lines (\Ha, \Hb, \NII, \SII\ and \OIII) across the observed region of NGC~613. The emission line flux maps show the total flux (summed across all kinematic components) in the relevant emission line, for each spaxel in which the relevant line is detected with \mbox{S/N $>$ 5}. The black asterisks mark the peak of the integrated continuum emission, assumed to be the galaxy centre. The black contours indicate the regions containing 68 per cent (1$\sigma$) and 95 per cent (2$\sigma$) of the flux in each image. 

The integrated continuum emission and the recombination line (\Ha\ and \Hb) emission show similar spatial distributions but with a 2 arcsec ($\sim$260~pc) offset between their peaks. The integrated continuum emission primarily traces emission from older stellar populations, and decreases smoothly with galactocentric distance but is elongated in the \mbox{NW -- SE} direction (along the projected direction of the galaxy bar). On the other hand, the recombination line emission is expected to be excited primarily by ionizing photons from young, massive stars. The recombination line emission is also elongated in the \mbox{NW -- SE} direction, but peaks to the south of the galaxy nucleus, suggesting that the majority of the current star formation is occurring in the circumnuclear regions.  

\begin{figure*}
\centerline{\includegraphics[scale = 1.0, clip = true, trim = 0 235 0 0]{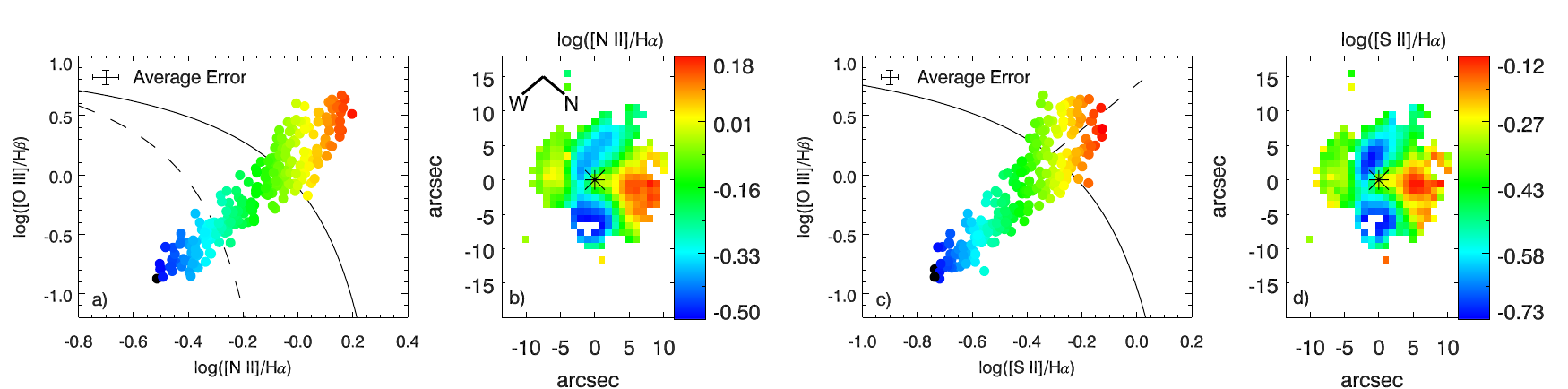}}
\caption{Panels a) and c): \NIIHa\ and \SIIHa\ vs. \OIIIHb\ diagnostic diagrams for NGC~613 populated with line ratios calculated from the total emission line fluxes (summed across all kinematic components) of individual spaxels across the integral field datacube. In these and all subsequent diagnostic diagrams we only include spectra for which all five diagnostics lines are detected to at least the 5$\sigma$ level. The solid black curves trace the \citet{Ke01a} theoretical upper bound to pure star formation. The dashed black curves on the \NIIHa\ and \SIIHa\ vs. \OIIIHb\ diagnostic diagrams delineate the \citet{Ka03} and \citet{Ke06} empirical classification lines (Ka03 and Ke06 lines), respectively. The data are colour-coded by the line ratio on the $x$-axis of each diagnostic diagram. Panels b) and d): maps of the \NIIHa\ and \SIIHa\ ratios across NGC~613, using the same colour-coding. The black asterisks mark the galaxy centre. Only spaxels in which all five diagnostic lines are detected are included in these line ratio maps.}
\label{fig:line_ratio_maps}
\end{figure*}

The \OIII\ emission peaks 1.4 arcsec ($\sim$180~pc) to the north of the galaxy centre and is elongated in the NE -- SW direction - approximately perpendicular to the elongation observed in the recombination line emission, and aligned with the inner radio jet structure detected by \citet{Hummel87}. The different spatial distributions of the recombination line and \OIII\ emission suggest that multiple ionization mechanisms are likely to contribute significantly to the observed line emission across the central region of NGC~613. The recombination line strengths are primarily determined by the atomic gas density, whereas the forbidden line strengths are strongly dependent on the local collisional excitation rate and are therefore sensitive to the presence of a hard ionizing radiation field and/or shocks. The stellar ionizing radiation field may therefore be the dominant ionization mechanism to the NW and SE of the galaxy nucleus (as traced by the recombination line emission), whereas shock excitation and/or AGN activity may be responsible for ionizing the majority of the gas to the NE and SW of the nucleus (as traced by the \OIII\ emission). 

\subsubsection{Diagnostic Diagrams and Line Ratio Maps}
\label{subsubsec:diagnostic_diagrams}
The relative contributions of different ionization mechanisms to the line emission across the nuclear region of NGC~613 can be probed quantitatively using line ratios. Panels a) and c) of Figure \ref{fig:line_ratio_maps} show the \NIIHa\ and \SIIHa\ vs. \OIIIHb\ diagnostic diagrams for the central region of NGC~613, where each solid circle represents the line ratios calculated from the total emission line fluxes (summed across all kinematic components) of an individual spaxel. In these and all subsequent diagnostic diagrams we only include spectra for which all five diagnostic lines are detected to at least the 5$\sigma$ level. The data are colour-coded according to the ratio on the $x$-axis of the diagnostic diagram. 

The diagnostic power of the emission line ratios allows for the use of classification lines to separate spectra dominated by different ionization mechanisms. The solid black curves on both diagnostic diagrams trace the \citet{Ke01a} theoretical upper bound to pure star formation (Ke01 line). All spectra lying above the Ke01 line are dominated by ionization mechanisms more energetic than star formation. The dashed black curve on the \NIIHa\ vs. \OIIIHb\ diagnostic diagram delineates the \citet{Ka03} empirical classification line (Ka03 line), which traces the upper boundary of the Sloan Digital Sky Survey (SDSS) star formation sequence. All spectra lying below the Ka03 line are dominated by star formation. Spectra lying between the Ka03 and Ke01 lines have significant contributions from both star formation and more energetic ionization mechanisms. The dashed black curve on the \SIIHa\ vs. \OIIIHb\ diagnostic diagram traces the \citet{Ke06} empirical classification line (Ke06 line), which separates high ionization spectra associated with Seyfert AGN activity (above the Ke06 line) from spectra characteristic of Low Ionization (Nuclear) Emission Regions (LI(N)ERs, below the Ke06 line), which could be associated with a variety of ionization mechanisms including shock excitation \citep[e.g.][]{Heckman80, Dopita95, Lipari04, Monreal-Ibero06, Rich11, Ho14}, low luminosity AGN (LLAGN) activity \citep[e.g.][]{Storchi-Bergmann97, Eracleous01, Ulvestad01, Ho01}, and post asymptotic giant branch (pAGB) stars \citep[e.g.][]{Binette94, Kehrig12, Belfiore16}. The \NIIHa, \SIIHa\ and \OIIIHb\ ratios enable the robust separation of spectra dominated by star formation, Seyfert AGN activity and shock excitation.

The line ratios of individual spaxels in NGC~613 fall along a tight line ratio sequence spanning from the star-forming region to the AGN/shock/pAGB region of the diagnostic diagrams. This line ratio sequence is likely to be driven by spatial variations in the fraction of emission associated with star formation and the fraction of emission associated with more energetic ionization mechanisms. Panels b) and d) of Figure \ref{fig:line_ratio_maps} show maps of the \NIIHa\ and \SIIHa\ ratios. Only spaxels in which all five diagnostic lines are detected are included in these line ratio maps. The smallest line ratios are observed to the NW and SE of the galaxy nucleus, consistent with our hypothesis that these regions are dominated by star formation (see Section \ref{subsubsec:flux_maps}). The largest line ratios are observed to the NE of the nucleus where we observe a strong excess in the \OIII\ emission relative to the recombination line emission (see Figure \ref{fig:flux_maps}), consistent with this region being dominated by more energetic ionization mechanisms. We also observe a small enhancement in the line ratios to the SW of the nucleus. The nucleus itself and the regions to the N, E, S and W of the nucleus have intermediate line ratios, suggestive of mixing between spectra associated with multiple ionization mechanisms.

\begin{figure*}
\centerline{\includegraphics[scale = 1, clip = true, trim = 10 180 0 0]{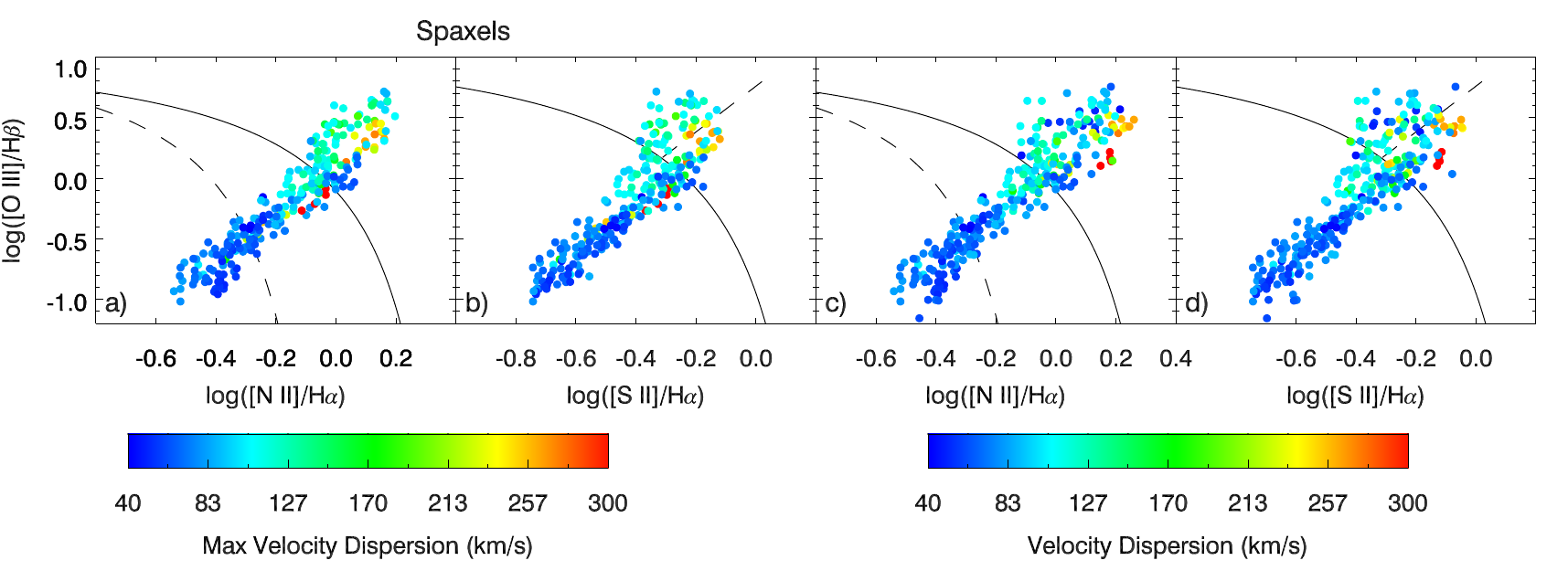}}
\caption{\NIIHa\ and \SIIHa\ vs. \OIIIHb\ diagnostic diagrams for NGC~613 with line ratios extracted from (a and b) the total emission line fluxes of individual spaxels, and (c and d) the spectra of individual kinematic components of individual spaxels across the integral field datacube. The data are colour-coded by (a and b) the maximum ionized gas velocity dispersion of any component in that spaxel, and (c and d) the ionized gas velocity dispersion measured for that component. The solid and dashed black curves are the same as in Figure \ref{fig:line_ratio_maps}.}
\label{fig:BPT_components}
\end{figure*}

The diagnostic diagrams and line ratio maps provide us with a clearer picture of how the fraction of emission associated with star formation varies across the nuclear region of NGC~613, but do not assist us in determining which ionization mechanism(s) are responsible for enhancing the forbidden line strengths to the NE and SW of the nucleus. The spaxels with the largest diagnostic line ratios lie in the transition region between Seyfert and LINER-like ionization on the \SIIHa\ vs. \OIIIHb\ diagnostic diagram, and thus could plausibly be consistent with Seyfert and/or LINER-like emission. The spatial elongation in the \OIII\ emission is approximately perpendicular to the elongation in the \Ha\ emission and is aligned with the inner radio jet. The morphology of the \OIII\ emission is therefore consistent with either an AGN ionization cone, or with an outflow in the jet direction producing shocks. Further information is required to constrain the ionization mechanisms responsible for enhancing the collisional excitation rate to the NE and SW of the nucleus in NGC~613.

\subsubsection{Kinematics}
Shock fronts are characterised by large velocity gradients, and therefore velocity dispersion is a key parameter for identifying shocks in galaxy spectra \citep[e.g.][]{Rich11, Ho14}.  Panels a) and b) of Figure \ref{fig:BPT_components} show the same diagnostic diagrams shown in Figure \ref{fig:line_ratio_maps}, but with the spaxels colour-coded according to the maximum velocity dispersion measured for any of the individual kinematic components in that spaxel. The spaxels with the largest maximum velocity dispersions preferentially lie on the under side of the mixing sequence (with a lower \OIIIHb\ ratio at a given \NIIHa\ ratio) and in the LINER region of the \SIIHa\ vs. \OIIIHb\ diagnostic diagram, suggesting that these spaxels may have significant contributions from shocks. We investigate this further by placing the line ratios of individual kinematic components on the diagnostic diagrams, colour-coded by the velocity dispersion measured for each component (panels c and d). The components in the LINER region of the \SIIHa\ vs. \OIIIHb\ diagnostic diagram have a larger average velocity dispersion than the components in the Seyfert region of the diagnostic diagram (155$\pm$14 km~s$^{-1}$ compared to 105$\pm$7 km~s$^{-1}$), and all the spaxels with $\sigma \ga$ 250 km s$^{-1}$ lie in the LINER region of the diagnostic diagram. The combination of LINER-like line ratios and high ionized gas velocity dispersion is indicative of shock excitation. The lower velocity dispersion components in the Seyfert region of the diagnostic diagram are likely to be ionized by the AGN radiation field. We therefore conclude that star formation, shock excitation and AGN activity all contribute significantly to the observed spectra in the nuclear region of NGC~613.

\section{Separating Star Formation, Shock Excitation and AGN Activity}
\label{sec:decomp}
In Section \ref{subsec:ionization_mechanisms}, we showed that the variations in line ratios across the nuclear region of NGC~613 can be explained by mixing between emission associated with star formation, shock excitation and AGN activity. Here, we build on this result and suggest that the extinction corrected emission line fluxes (line luminosities) of the observed spectra are well represented by linear superpositions of the line luminosities of a single set of three representative basis spectra - an \HII\ region spectrum, an AGN NLR spectrum and a shock excited spectrum. To test this scenario, we first describe how to select these basis spectra, and then use the technique presented by \citet{Davies16decomp} to separate the line luminosities of the observed spectra into contributions from star formation, shock excitation and AGN activity. Finally, we compare the derived star formation, shock and AGN components to independent tracers of these ionization mechanisms at other wavelengths.

\begin{figure*}
\centerline{\includegraphics[scale=1, clip = true, trim = 0 165 150 40]{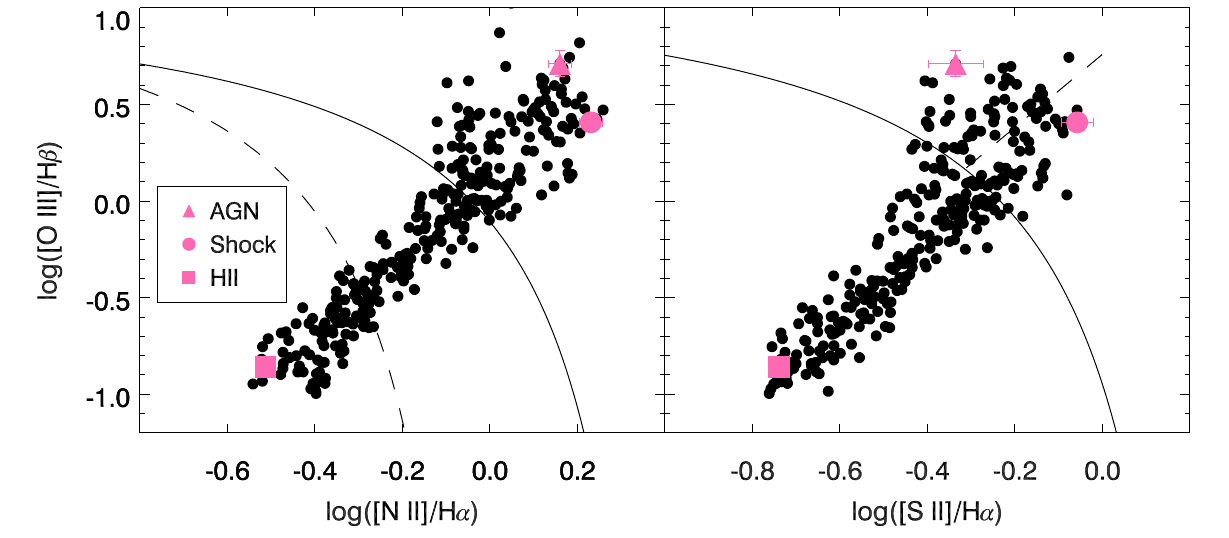}}
\caption{\NIIHa\ and \SIIHa\ vs. \OIIIHb\ diagnostic diagrams populated with line ratios extracted from the emission line fluxes of individual kinematic components within individual spaxels (black points). The pink squares, triangles and circles indicate the line ratios of the basis spectra selected to represent pure star formation, AGN activity and shock excitation, respectively. The solid and dashed black curves are the same as in Figure \ref{fig:line_ratio_maps}.}
\label{fig:basis_spectra}
\end{figure*}

\subsection{Selection of Basis Spectra}
\label{subsec:basis_spectra}
The S7 spectra of NGC~613 show evidence for ionization by star formation, shock excitation and AGN activity (see Section \ref{subsec:ionization_mechanisms}), and therefore we require one basis spectrum associated with each of these ionization mechanisms. We empirically select basis spectra as illustrated in Figure \ref{fig:basis_spectra}. The diagnostic diagrams are populated with line ratios calculated from the emission line fluxes of individual kinematic components within individual spaxels. Multiple kinematic components arise when emission associated with several different physical processes is superimposed along the line of sight. Therefore, selecting the basis spectra using line ratios extracted from the fluxes of individual kinematic components of individual spaxels minimises the level of contamination from emission associated with other ionization mechanisms.

The pink squares, triangles and circles indicate the line ratios of the basis spectra chosen to represent pure star formation, AGN activity and shock excitation, respectively. Each basis spectrum has line ratios consistent with ionization by a single mechanism.

We note that the basis spectra themselves are each superpositions of many spectra associated with the relevant ionization mechanism. The spatial resolution of our observations ($\sim$190~pc) is an order of magnitude or more larger than the typical size of an \HII\ region, and therefore the \HII\ region basis spectrum is likely to be a luminosity-weighted mean of many individual \HII\ region spectra. Similarly, the AGN NLR and shock excited basis spectra are likely to be superpositions of emission associated with multiple gas clouds superimposed along the line of sight.

\subsection{Decomposition Method}
\label{subsec:method}
We assume that the luminosity $L$ of any emission line $i$ in the spectrum of any spaxel $j$ from the integral field datacube of NGC~613 can be expressed as a linear superposition of the line luminosities of the \HII\ region basis spectrum, the shock excited basis spectrum and the AGN NLR basis spectrum: 
\begin{equation}
L_{i}(j) = m(j) \times L_{i}(\textrm{HII})~+~n(j) \times L_{i}(\textrm{AGN})~+~k(j) \times L_{i}(\textrm{shock})
\label{eqn:superposition}
\end{equation}
Here $L_{i}(\rm{HII})$, $L_{i}(\rm{AGN})$ and $L_{i}(\rm{shock})$ are the luminosities of the \HII\ region, AGN NLR, and shock excited basis spectra, respectively, in emission line $i$. The coefficient $m(j)$ is the ratio of the SFR calculated from spectrum $j$ to the SFR calculated from the \HII\ region basis spectrum. Similarly, $n(j)$ and $k(j)$ are the ratios of the AGN and shock luminosities calculated from spectrum $j$ to the luminosities calculated from the AGN NLR and shock excited basis spectra, respectively. The superposition coefficients $m(j)$, $n(j)$ and $k(j)$ vary between spectra but are the same for all emission lines within a given spectrum. 

We use Equation \ref{eqn:superposition} and the basis spectra described in Section \ref{subsec:basis_spectra} to separate the total emission line luminosities (summed across all kinematic components) of individual spaxels of NGC~613 into contributions from star formation, shock excitation and AGN activity. Prior to decomposition, we correct the emission line fluxes of all spectra (the basis spectra and the spectra of individual spaxels) for extinction using the Balmer decrement, assuming an intrinsic \Ha/\Hb\ ratio of 2.86 and adopting the \citet{Fischera05} extinction curve with $R_V^A$~=~4.5. The intrinsic \Ha/\Hb\ ratio is known to be higher in AGN NLRs ($\sim$3.1; \citealt{Osterbrock06}), but we do not know \textit{a priori} which spaxels are AGN dominated and therefore we choose to adopt a single intrinsic Balmer decrement. In reality this choice has very little impact on the corrected line luminosities (a factor of 1.35 for \OIII) and therefore does not have a significant impact on our results.

We perform the decomposition using the total line luminosities of individual spaxels (rather than the luminosities of individual kinematic components within those spaxels) to maximise the S/N on each emission line. We only perform the decomposition on spaxels for which all five diagnostic lines are detected to at least the 5$\sigma$ level (where the S/N is calculated for the flux summed across all kinematic components). For each spaxel, we calculate the superposition coefficients by performing least squares minimisation on Equation \ref{eqn:superposition} applied to the extinction-corrected fluxes of the four strongest emission lines in our data (\Ha, \NII, \SII\ and \OIII). The three basis spectra are normalized to an \Ha\ luminosity of 1 to ensure that the least squares minimisation is primarily sensitive to the relative (rather than absolute) fluxes of the emission lines. The extinction correction fixes the \Ha/\Hb\ ratio and therefore including \Hb\ in the minimisation would not provide any additional constraints on the superposition coefficients. 

\section{Results}
\label{sec:results}  

\begin{figure*}
\centerline{\includegraphics[scale = 1, clip = true, trim = 0 15 120 0]{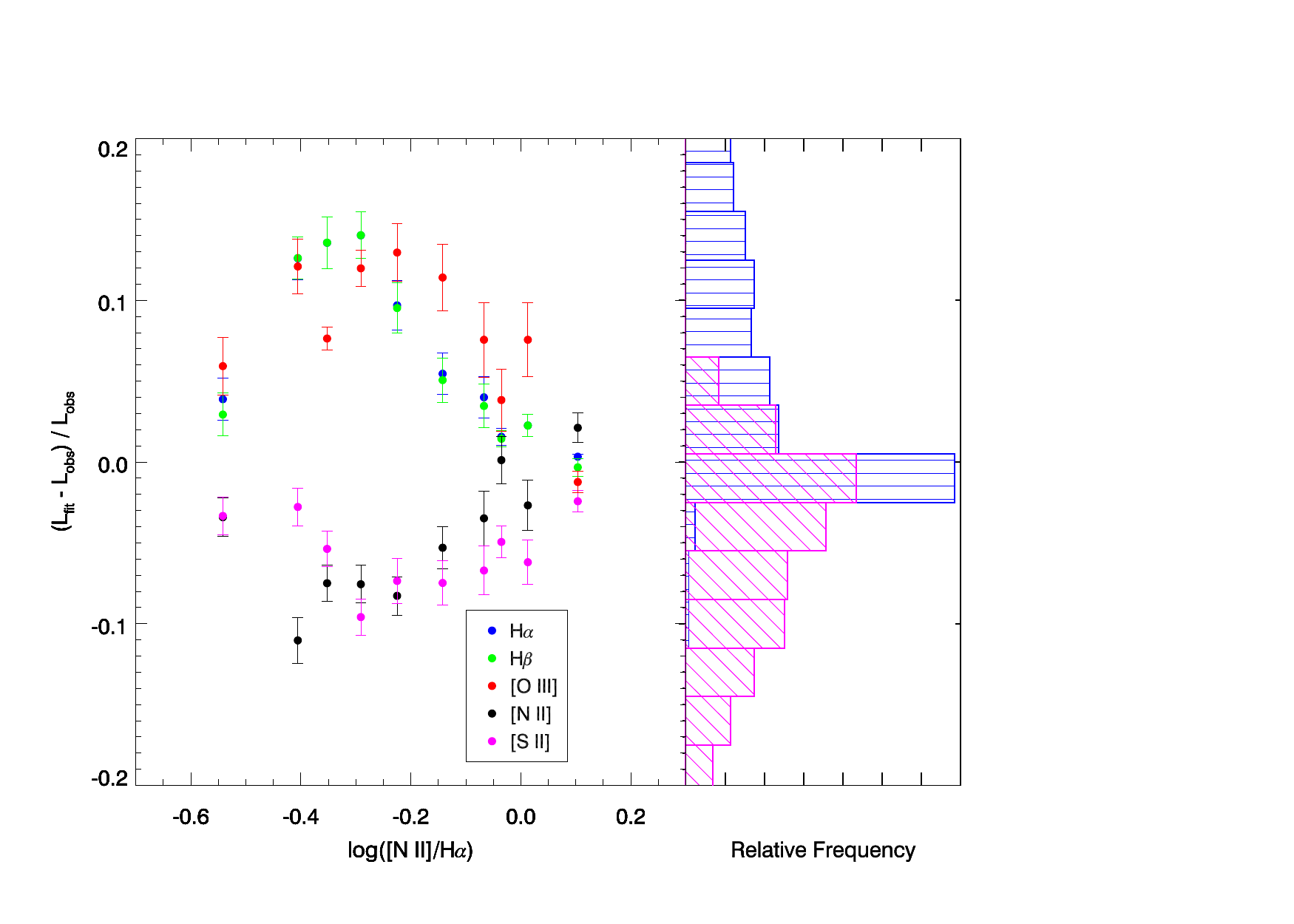}}
\caption{(Left) Relative difference between the fit and measured luminosities for each of the strong emission lines, as a function of the \NIIHa\ ratio. The data points and error bars represent the average value and standard deviation for pixels in each line ratio bin. (Right) Histograms of the relative difference between the fit and measured luminosities for the \Ha\ (blue) and \SII\ (magenta) lines, for all pixels.}
\label{fig:residuals}
\end{figure*}

\subsection{Accuracy of Recovered Luminosities}
\label{subsec:residuals}
We use the superposition coefficients ($m(j)$, $n(j)$ and $k(j)$) derived as described in Section \ref{subsec:method} to calculate the \Ha, \Hb, \NII, \SII\ and \OIII\ luminosities associated with star formation, shock excitation, and AGN activity for each spaxel across NGC~613. We find that the luminosities of individual emission lines in the spectra of individual spaxels across the datacube of NGC~613 are well reproduced by linear superpositions of the line luminosities extracted from the \HII\ region, AGN NLR and shock excited basis spectra. The left hand panel of Figure \ref{fig:residuals} shows how the relative difference between the fit and measured luminosities for each emission line varies as a function of the \NIIHa\ ratio. The pixels are divided into ten bins based on their \NIIHa\ ratios, and the plotted data points and error bars represent the average value and standard deviation for the pixels in each bin. We see that on average the fit luminosities do not deviate from the measured luminosities by more than 15 per cent, regardless of the \NIIHa\ ratio. The right hand panel of Figure \ref{fig:residuals} shows the distributions of the relative differences between the fit and measured luminosities for the \Ha\ and \SII\ lines, for all pixels. In more than 99 per cent of pixels, the fit luminosities match the measured luminosities to within 20 per cent.

It is interesting to note that the \Ha, \Hb\ and \OIII\ luminosities are systematically over-estimated whereas the \NII\ and \SII\ luminosities are systematically under-estimated. In addition, the magnitude of the relative difference between the fit and measured luminosities increases with decreasing \NIIHa\ ratio, except at the very lowest \NIIHa\ ratio where the magnitude of the relative difference is less than 5 per cent. These features may indicate that the \NIIHa\ and \SIIHa\ ratios of the \HII\ region basis spectrum are slightly under-estimated.

\begin{figure*}
\centerline{\includegraphics[scale=0.95, clip = true, trim = 0 15 0 30]{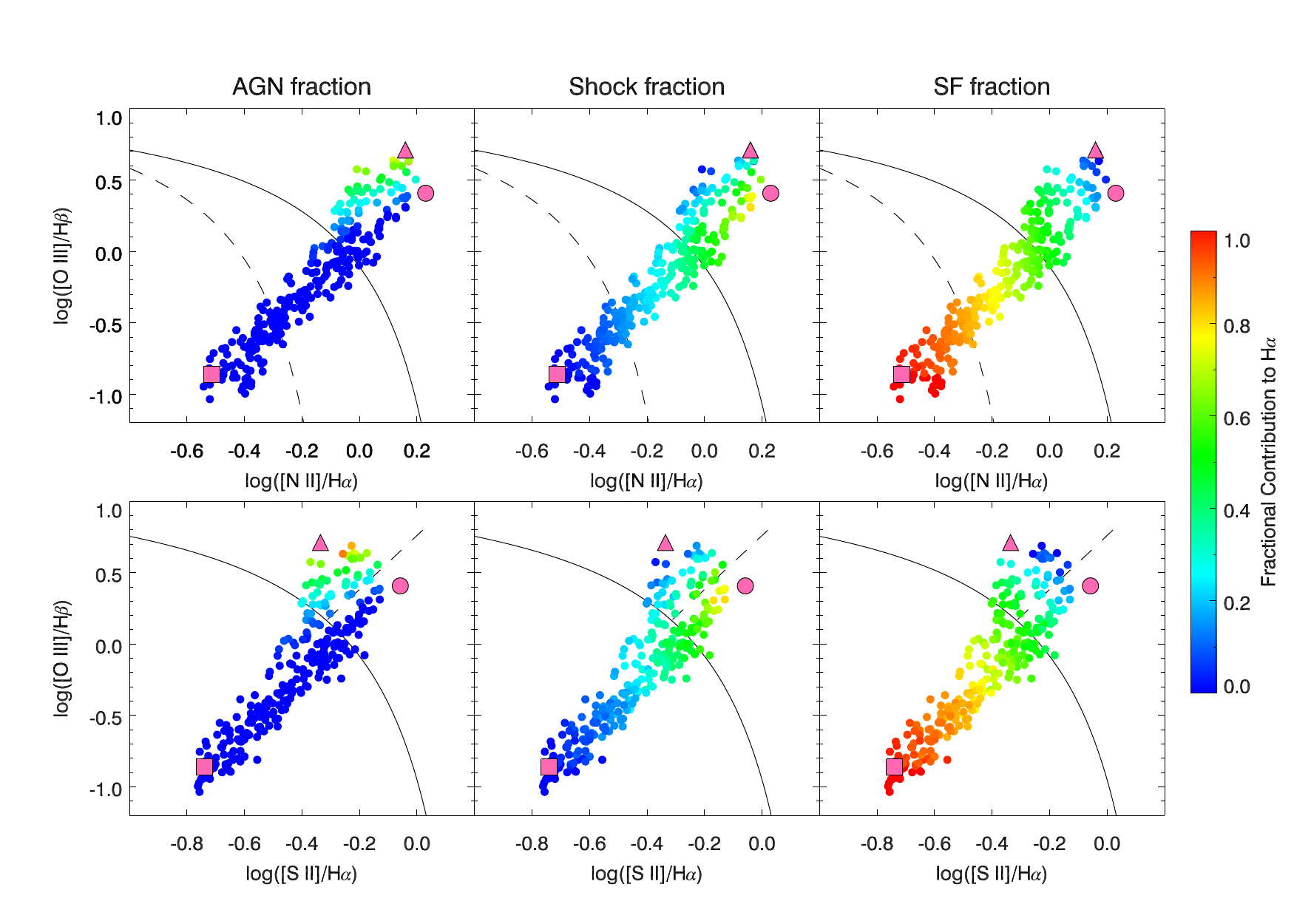}}
\caption{\NIIHa\ (top) and \SIIHa\ (bottom) vs. \OIIIHb\ diagnostic diagrams for NGC~613 with line ratios extracted from total emission line fluxes of individual spaxels. The \HII\ region, AGN NLR and shock basis spectra are indicated by pink squares, triangles and circles, respectively. The spectra are colour-coded by the fraction of \Ha\ emission attributable to AGN activity (left panels), shock excitation (middle panels) and star formation (right panels), as determined by our spectral decomposition.}
\label{fig:diagnostic_diagrams}
\end{figure*} 

\subsection{Fractional Contributions of Ionization Mechanisms as a Function of Diagnostic Line Ratios}
The most fundamental test of our decomposition method is to check that the fraction of emission attributed to each ionization mechanism varies smoothly as a function of the diagnostic line ratios, peaking at the line ratios of the relevant basis spectrum and decreasing as the line ratios become closer to those of the other basis spectra. Figure \ref{fig:diagnostic_diagrams} shows the \NIIHa\ (top row) and \SIIHa\ (bottom) vs. \OIIIHb\ diagnostic diagrams for NGC~613, where each solid circle represents the line ratios calculated from the total emission line fluxes of an individual spaxel. The data are colour-coded according to the fraction of \Ha\ emission attributable to AGN activity (left panels), shock excitation (middle) and star formation (right). The relative contribution of each ionization mechanism to the \Ha\ luminosity evolves smoothly as a function of the diagnostic line ratios, indicating that the least squares fitting algorithm used to constrain the superposition coefficients behaves as expected. We note that many of the spaxels in the star forming region of the \SIIHa\ vs. \OIIIHb\ diagnostic diagram (below the Ke01 line) have a non-negligible contribution from shocks. This is a known feature of the \SIIHa\ vs. \OIIIHb\ diagnostic diagram, which does not differentiate between star formation dominated and composite spectra \citep[see][]{Ke06}.

Our results indicate that even above the \citet{Ke01a} line, in the AGN region of the \NIIHa\ vs. \OIIIHb\ diagnostic diagram, star formation can still be responsible for over 50\% of the \Ha\ emission. The \citet{Ke01a} line delineates the boundary beyond which the line ratios cannot be explained by \textit{pure} star formation, but star formation can still be the \textit{dominant} source of \Ha\ emission in this regime. We emphasise that the contribution of star formation to spectra lying above the \citet{Ke01a} line should not be neglected (see also \citealt{CidFernandes11}).

\subsection{Mapping Emission Associated with Different Ionization Mechanisms across NGC~613}
\begin{figure*}
\centerline{\includegraphics[scale=1.1]{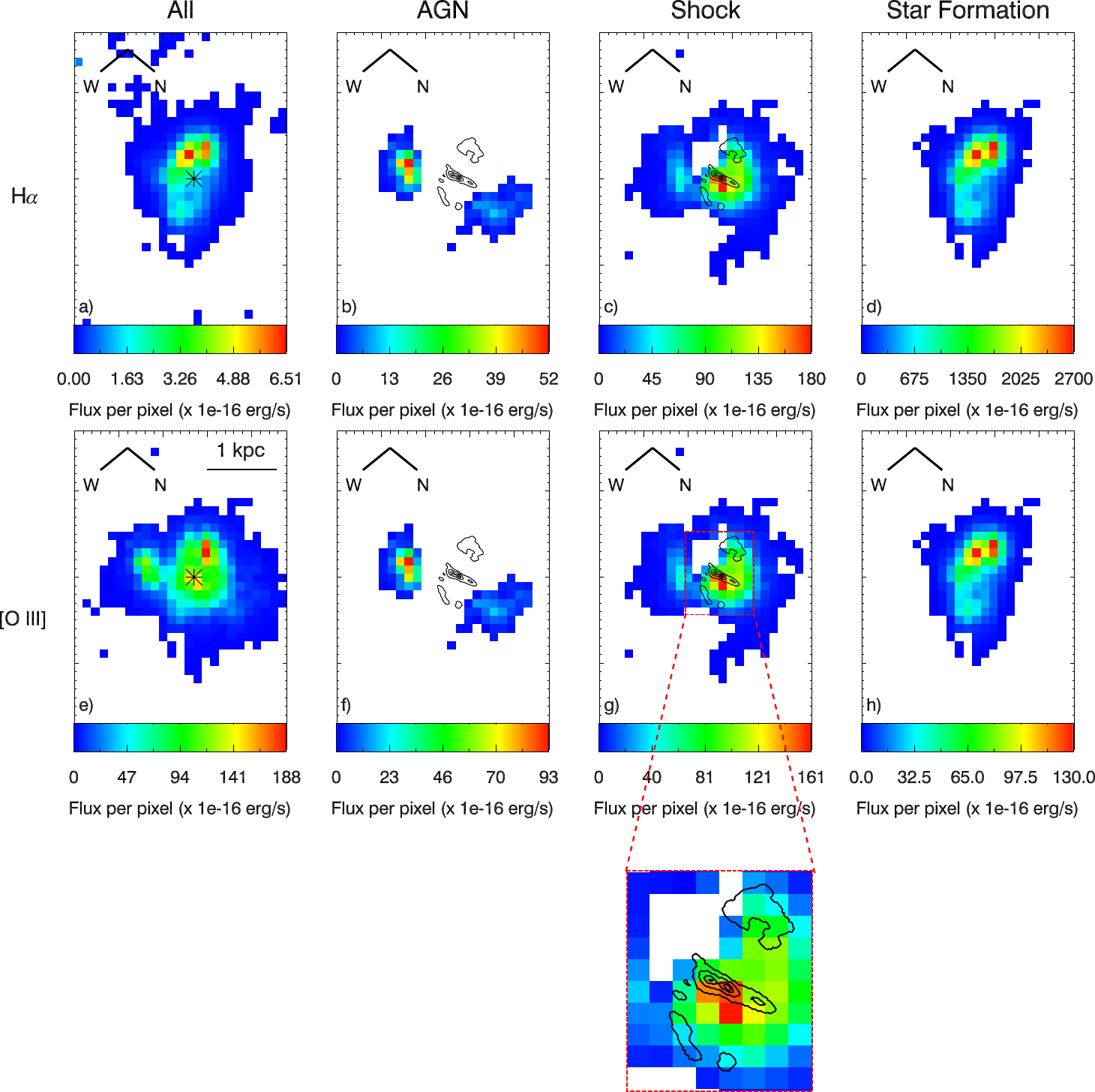}}
\caption{Maps of the extinction-corrected \Ha\ and \OIII\ emission across NGC~613 (a and e), the \Ha\ and \OIII\ emission attributable to AGN activity (b and f), shock excitation (c and g) and star formation (d and h). The angular resolution of the emission line images is 1.5 arcsec. The black contours in panels b), c), f) and g) trace the 4.89 GHz emission from the VLA archival dataset AH0231 \citep{Hummel92}. The contours represent levels of 7, 30, 70 and 90 per cent of the peak surface brightness of 2.33 mJy beam$^{-1}$, for a beam size of 0.54 arcsec (about half the size of a WiFeS pixel). The 4.89 GHz emission traces a compact nuclear jet structure and a central star forming ring. The black asterisks in panels a) and e) mark the galaxy centre.}
\label{fig:source_distribution}
\end{figure*}

Figure \ref{fig:source_distribution} shows the spatial distributions of the total \Ha\ and \OIII\ emission (panels a and e) and the \Ha\ and \OIII\ emission associated with star formation, shock excitation, and AGN activity (panels b-d and f-h) in each spaxel across the central region of NGC~613. We evaluate the success of our decomposition by comparing the decomposed emission line maps with maps of the \mbox{4.89 GHz} flux density and the 4.6-8.1 GHz spectral index. The contours overlaid in panels b), c), f) and g) of Figure \ref{fig:source_distribution} trace the 4.89~GHz emission detected by the \textit{Very Large Array} (VLA) in the A configuration (project AH0231, PI: K. Hummel). For this paper we have re-imaged the archival data, originally published by \citet{Hummel92}, to create a new 4.89~GHz image with a beam size of 0.54 arcsec (approximately half the size of a WiFeS pixel). The contours trace the ridge lines for emission at levels of 7, 30, 70 and 90 per cent of the peak surface brightness of \mbox{2.33 mJy beam$^{-1}$}.

The 4.89 GHz emission appears to have a composite structure, consisting of a nuclear jet and a star forming ring (seen previously in the radio by \citealt{Hummel87} and \citealt{Hummel92} and in the infrared by \citealt{FalconBarroso14}). We further investigate the physical origins of these radio structures by using \textit{Australia Telescope Compact Array} \citep[ATCA;][]{Frater92} data to construct a map of the 4.6-8.1 GHz spectral index. (A description of the ATCA observations, data reduction and processing can be found in Appendix A). Figure \ref{fig:spectral_index} shows the spectral index map with contours of the VLA 4.89~GHz emission overlaid in black. We find clear regions of flatter spectral index along the eastern and western ridges of the image, close to the star-forming ring structure that is observed in the VLA image, although the western ridge is noisier. The mean spectral index value on the eastern side is \mbox{$\alpha$ = -0.1 $\pm$ 0.2} and the mean spectral index in the western side is \mbox{$\alpha$ = -0.2 $\pm$ 0.3}, consistent with free-free emission from \HII\ regions. On the other hand, the mean spectral index along the jet is \mbox{$\alpha$ = -0.96 $\pm$ 0.09}, consistent with optically thin synchrotron jet emission. The ATCA data therefore confirm the presence of the star-forming ring and nuclear jet structure seen in the VLA data.

In the following sections we use the high spatial resolution VLA radio images and the decomposed WiFeS emission line maps to infer the physical origin and luminosity of the emission associated with star formation, shock excitation and AGN activity in the central region of NGC~613.

\subsubsection{AGN Activity}
Panels b) and f) of Figure \ref{fig:source_distribution} show maps of the \Ha\ and \OIII\ emission associated with AGN activity. The optical line emission excited by the AGN ionizing radiation field is distributed in two clumps on opposite sides of the galaxy nucleus. The clumps are aligned with the nuclear radio jet structure, and are therefore likely to trace parts of an AGN ionization cone. This ionization cone is thought to consist of material entrained in an AGN driven outflow \citep[see e.g.][]{Condon87, Hummel87, Hummel92}.

The \OIII\ luminosity attributable to AGN accretion is \mbox{(7.34$\pm$0.93)$\times$10$^{39}$ erg s$^{-1}$}, which corresponds to an AGN bolometric luminosity of \mbox{$L_{bol, AGN}$ = 4.0$\times$10$^{42}$ erg s$^{-1}$} (using a bolometric correction factor of 600; \citealt{Kauffmann09}). We test the accuracy of this bolometric luminosity estimate by comparing it to the AGN bolometric luminosity calculated from the hard X-ray (2-10~keV) emission, of which the vast majority is expected to be associated with AGN activity. NGC~613 has a \mbox{2-10 keV} luminosity of \mbox{8.92$\times$10$^{40}$ erg s$^{-1}$} (using the flux measured by \citealt{Castangia13} and a distance of 26.4 Mpc), which corresponds to an AGN bolometric luminosity of \mbox{$L_{bol, AGN}$ = 1.6$\times$10$^{42}$ erg s$^{-1}$} (using an average bolometric correction factor of 20; \citealt{Vasudevan10}). In comparison, the AGN bolometric luminosity calculated from the total \OIII\ emission (including contributions from shock excitation and star formation) is \mbox{$L_{bol, [O III]}$ = 3.75$\times$10$^{43}$ erg s$^{-1}$}. Even limiting this to the central \mbox{$1\times1$ kpc$^{2}$}, the AGN emission still only accounts for $\sim$10\% of the total [OIII] luminosity (\mbox{$L_{bol, [O III]}$ = 3.29$\times$ 10$^{43}$ erg $s^{−1}$)}. Correcting the \OIII\ emission for the contribution of star formation and shock excitation significantly improves the accuracy of the AGN bolometric luminosity calculated from the \OIII\ emission.

\begin{figure}
\centerline{\includegraphics[scale=0.6, clip = true, trim = 0 0 0 6]{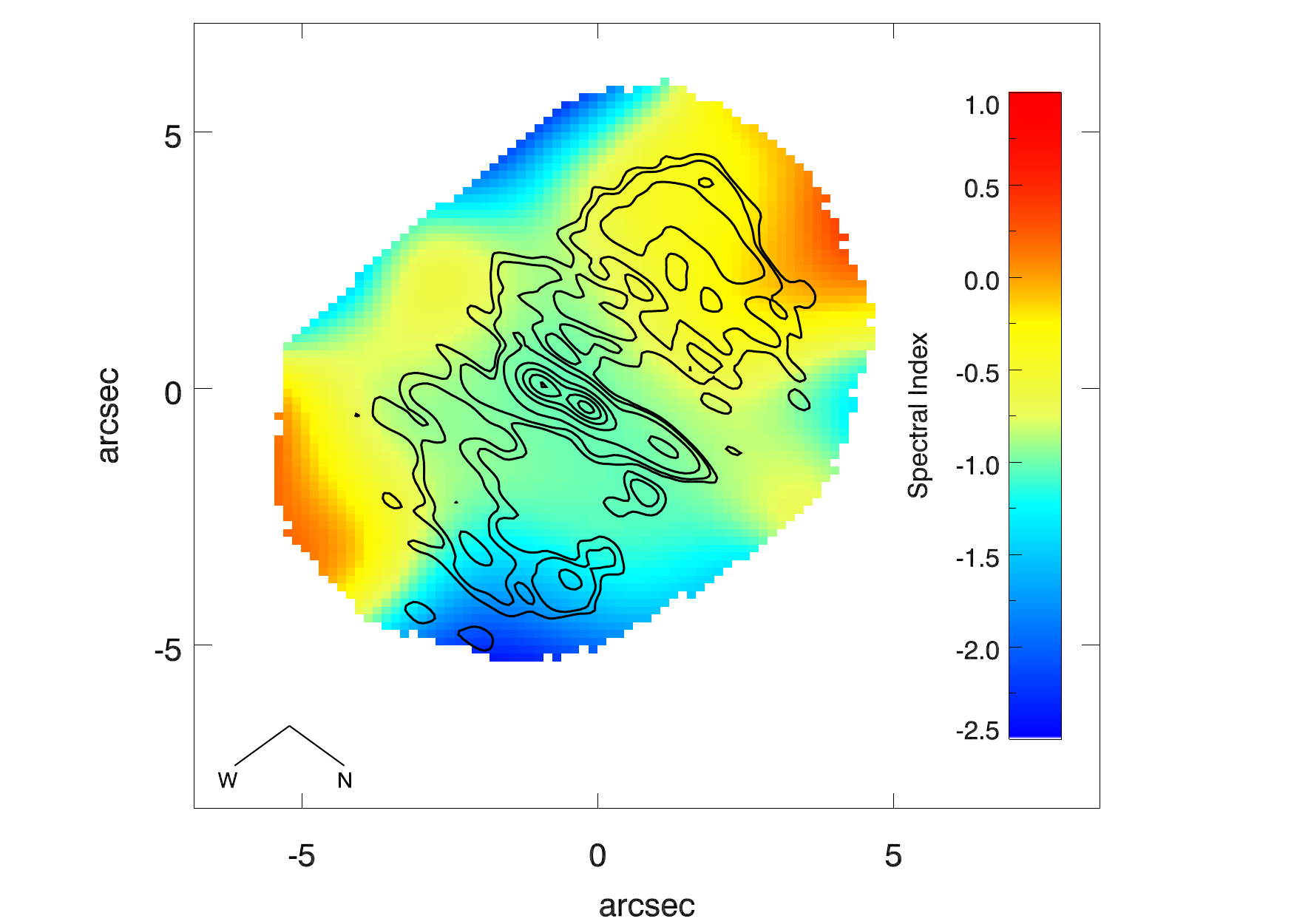}}
\caption{Map of the 4.6-8.1 GHz spectral index, constructed from ATCA data. The overlaid contours trace the 4.89 GHz emission detected by the VLA. The contour at levels of 3.5, 5, 10, 30, 50, 70 and 90 per cent of the peak surface brightness.}
\label{fig:spectral_index}
\end{figure}

\subsubsection{Shock Excitation}
\begin{figure}
\centerline{\includegraphics[scale=1.1, clip = true, trim = 220 165 65 40]{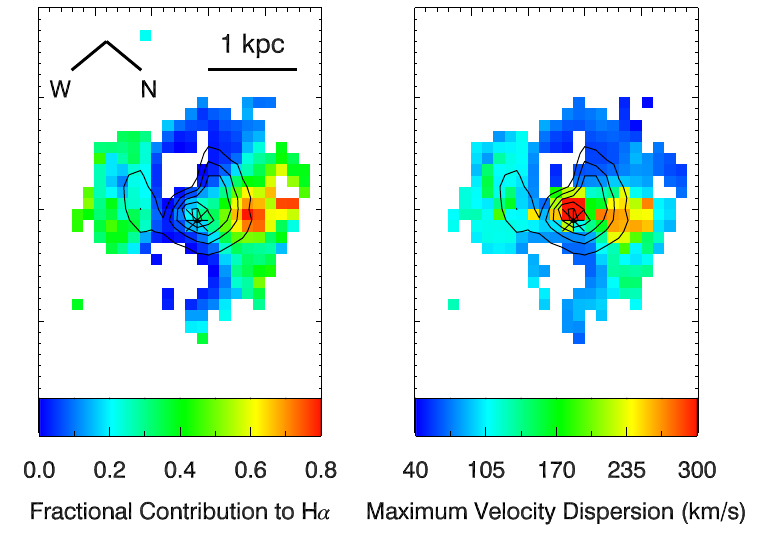}}
\caption{Maps of the fractional contribution of shocks to the \Ha\ emission (left) and the maximum velocity dispersion measured for any of the kinematic components (right) across NGC~613. The black contours trace the \Ha\ emission associated with shock excitation (as seen in panel c of Figure \ref{fig:source_distribution}) at levels of 10, 30, 50, 70 and 90 per cent of the peak luminosity. The black asterisks indicate the peak of the shock emission}.
\label{fig:fshock}
\end{figure}

Panels c) and g) of Figure \ref{fig:source_distribution} show maps of the \Ha\ and \OIII\ emission associated with shock excitation. The peak of the optical line emission associated with shock excitation is coincident with the peak of the compact jet structure seen at 4.89~GHz. Figure \ref{fig:fshock} shows maps of the fractional contribution of shock excitation to the \Ha\ emission in each spaxel (left), and the maximum velocity dispersion measured for any of the kinematic components in each spaxel (right). Black contours trace the \Ha\ emission associated with shock excitation (as seen in panel c of Figure \ref{fig:source_distribution}) at levels of 10, 30, 50, 70 and 90 per cent of the peak luminosity. Black asterisks indicate the location of the shock emission peak.

The shock emission peak is characterised by a high ionized gas velocity dispersion (\mbox{$\sigma\ \ga$ 300 km s$^{-1}$}), and the fractional contribution of shocks to the \Ha\ emission is relatively high (\mbox{$f_{\rm shock} \sim$ 30-50 per cent}). Strong $H_2$ and \mbox{[Fe II]} emission have been previously observed in the central \mbox{$\sim$1~$\times$~1 arcsec$^2$} region of NGC~613 \citep{FalconBarroso14}, providing independent confirmation of the presence of shocks. The velocity dispersion generally remains elevated (\mbox{$\sigma\ \ga$ 100 km s$^{-1}$}) to the NE and SW of the nucleus where the two lobes of the AGN ionization cone are observed. The shock fraction increases moving off the disk of the galaxy (peaking at $\sim$80 per cent to the NE of the nucleus) as the emission from \HII\ regions fades. The shock emission in the high shock fraction (\mbox{$f_{\rm shock} \ga$~30 per cent}) regions is likely to trace the outer boundary of the AGN ionization cone where outflowing material is interacting with the surrounding interstellar medium \citep{FalconBarroso14, Hummel87, Hummel92}. 

The \Ha\ luminosity associated with the high velocity dispersion (\mbox{$\sigma \ga$ 100 km s$^{-1}$}) shock excited material is \mbox{(4.00$\pm$0.19)$\times$10$^{40}$ erg s$^{-1}$}, corresponding to a shock luminosity of \mbox{$\sim$7-24$\times$10$^{41}$ erg s$^{-1}$} (depending on the shock velocity and the fraction of hydrogen that is pre-ionized; \citealt{Rich10}). The nuclear radio jet is transferring a significant amount of energy to the interstellar medium with which they are interacting, and this energy is being radiated away in the emission we observe.

\subsubsection{Star Formation}
Star formation is the dominant source of \Ha\ emission and a significant source of \OIII\ emission across the nuclear region of NGC~613. The peak of the line emission associated with star formation is coincident with the south eastern shock emission peak and is likely to be associated with the nuclear star forming ring.

\begin{figure}
\centerline{\includegraphics[scale=1, clip = true, trim = 40 175 240 35]{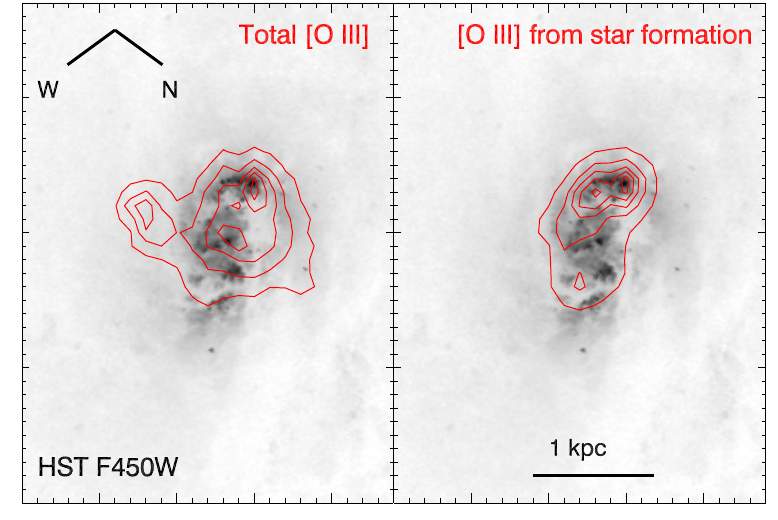}}
\caption{Zoom in on the central 25$\times$38 arcsec$^2$ region of an HST F450W image of NGC~613, with contours of the total \OIII\ emission (left) and \OIII\ emission due to star formation (right) overlaid in red. The contours are at levels of 10, 30, 50, 70 and 90 per cent of the peak \OIII\ emission in each panel.}
\label{fig:overlay}
\end{figure}

The line emission associated with star formation generally originates from the regions with the strongest stellar continuum emission. Figure \ref{fig:overlay} shows the central \mbox{25$\times$38 arcsec$^2$} region of a \emph{Hubble Space Telescope} (HST) F450W image of NGC~613, with contours of the total \OIII\ emission (left), and the \OIII\ emission due to star formation (right), overlaid in red. The contours are at levels of 10, 30, 50, 70 and 90 per cent of the peak \OIII\ emission in each panel. The F450W filter covers 358-533~nm and is expected to be dominated by stellar continuum in regions with active star formation. The total \OIII\ emission has a very different spatial distribution from that of the F450W emission, suggesting that ionization mechanism(s) other than star formation are likely to have a significant contribution to the total \OIII\ emission across the central region of NGC~613. In contrast, the \OIII\ emission associated with star formation is coincident with the regions of the strongest F450W emission. Our decomposition allows us to isolate the contribution of star formation to emission lines (such as \OIII) that are typically dominated by emission associated with harder ionization mechanisms.

The \Ha\ luminosity associated with star formation is \mbox{(5.28 $\pm$ 0.59)$\times$10$^{41}$ erg s$^{-1}$}, which corresponds to an SFR of \mbox{4.15 $\pm$ 0.47 M$_\odot$ yr$^{-1}$} \citep{Kennicutt98}. We compare this SFR to the SFR calculated from the 12$\mu$m emission over the same region. The 12$\mu$m Wide-field Infrared Survey Explorer (WISE) band (W3) covers a strong polycyclic aromatic hydrocarbon (PAH) feature at 11.3$\mu$m. PAHs are excited by radiation from young massive stars and therefore trace active star formation. We use convolution kernels from \citet{Aniano11} to convolve the image of the \Ha\ luminosity associated with star formation (\Ha$_{SF}$) to the same PSF as the 12$\mu$m image, and then calculate the total \Ha$_{SF}$ flux and the total 12$\mu$m flux density within the WiFeS FOV. We convert the \Ha$_{SF}$ and 12$\mu$m luminosities to SFRs using the calibrations of \citet{Kennicutt98} and \citet{Cluver14}, respectively. The SFR calculated from the convolved \Ha$_{SF}$ image is \mbox{2.88 $\pm$ 0.40 M$_\odot$ yr$^{-1}$}, in close agreement with the 12$\mu$m SFR of \mbox{2.63 $\pm$ 0.09 M$_\odot$ yr$^{-1}$}. The SFR calculated from the convolved \Ha$_{SF}$ image is lower than the SFR calculated from the unconvolved image because the smoothing moves flux outside the boundary of the WiFeS FOV. Star formation is the dominant source of \Ha\ emission in the nuclear region of NGC~613, and therefore the calculated SFR does not change significantly if the total \Ha\ emission is used rather than the \Ha\ emission associated with star formation.

\section{Caveats}
\label{sec:discussion}
The relative luminosities of the emission lines used in the decomposition are primarily determined by the relative contributions of different ionization mechanisms to the line emission, but are also sensitive to the metallicity and ionization parameter of the ionized gas \citep{Groves04, Allen08, Dopita13}. The spectra analysed in this paper are extracted from the central $\sim$~3~$\times$~3 kpc$^2$ region of NGC~613, over which the metallicity is not expected to vary by more than 0.05 dex \citep{Ho15}. However, integral field studies of AGN host galaxies over larger spatial regions could probe much wider ranges in the metallicity and/or ionization parameter, potentially increasing the scatter in the diagnostic line ratios, and leading to discrepancies between the measured line luminosities and the luminosities obtained from the best-fit linear superpositions of the basis spectra line luminosities. We emphasise that the results of the decomposition should only be used if the best-fit luminosities are consistent with the measured luminosities for the vast majority of spaxels. 

Dust attenuation can significantly impact the derived spatial distributions of the emission associated with each of the ionization mechanisms. Our decomposition method can only be applied to spectra for which all five diagnostic emission lines (\Ha, \Hb, \NII, \SII\ and \OIII) are detected at the 5$\sigma$ level. Strong dust attenuation in some regions of AGN host galaxies may inhibit the detection of some or all of these emission lines, and therefore prevent the underlying emission associated with star formation, shock excitation and/or AGN activity from being recovered. 

The spatial resolution of the observations directly determines the spatial resolution of the final decomposed emission line maps. Small changes in the spatial resolution can significantly impact our understanding of the physical origin of the emission associated with each of the ionization mechanisms. The NGC~613 data used in this paper have a relatively high spatial resolution of $\sim$190~pc, and any features in the emission line maps on physical scales larger than this can be robustly detected. We repeated the analysis presented in this paper on older, lower quality WiFeS observations of NGC~613 with a spatial resolution of 2.2 arcsec ($\sim$280~pc). The double peak in the shock emission map (seen in panels c and g of Figure \ref{fig:source_distribution}) is not recovered in the lower spatial resolution observations, making it impossible to distinguish between the supernova shocks in the vicinity of the nuclear star forming ring and the more energetic shocks associated with the AGN driven outflow.

High spatial and spectral resolution observations are key for selecting basis spectra with minimal contamination from emission associated with other ionization mechanisms. The higher the spatial resolution, the smaller the number of ionizing sources contributing to the line emission in each spaxel. The higher the spectral resolution, the more kinematic components it is possible to robustly detect in each spaxel and the lower the degree of mixing between emission associated with different ionization mechanisms within the spectrum of each kinematic component. Our red channel spectra of NGC~613 have a spectral resolution of R$\sim$7000 (43 km s$^{-1}$), which is only a factor of two to a few larger than the velocity dispersions typical of local \HII\ regions (e.g \citealt{Epinat10, Green14}). The high spectral resolution of our data allow us to separate emission associated with different physical processes across much of the central region of the galaxy. Reducing the spectral and/or spatial resolution of the NGC~613 observations would increase the degree of contamination in the basis spectra and therefore reduce the accuracy of the decomposition. We note that, despite the high spectral and spatial resolution of our data, the basis spectra used in the decomposition are very likely to have a small amount of contamination from emission associated with other ionization mechanisms.

\section{Summary and Conclusions}
\label{sec:conc}
In this paper we use optical integral field data to spatially and spectrally separate emission associated with star formation, shock excitation and AGN activity in the central region of the nearby \mbox{(D = 26.4 Mpc)} active barred spiral galaxy NGC~613. Previous optical, infrared and radio studies of NGC~613 have revealed extended star formation which is enhanced at the ends of the bar, enhanced nuclear emission and shocked gas in the circumnuclear regions. The radio spectral indices derived from our ATCA observations of NGC~613 are also consistent with a combination of synchrotron emission from a radio jet and free-free emission from \HII\ regions.

We use the distribution of optical spectra on the \NIIHa\ and \SIIHa\ vs. \OIIIHb\ diagnostic diagrams to select three `basis spectra' - one representative of pure star formation, one representative of pure AGN activity and one representative of pure shock excitation. We show that the \Ha, \Hb, \NII, \SII\ and \OIII\ luminosities of all spaxels (in which all five lines are detected with \mbox{S/N $>$ 5}) across the datacube of NGC~613 are consistent with linear superpositions of the line luminosities of the three basis spectra. 

We separate the luminosity of each diagnostic emission line in each spaxel into contributions from star formation, shock excitation and AGN activity, and compare our decomposed emission line maps to independent tracers of each ionization mechanism at other wavelengths. We find that:
\begin{itemize}
\item{The line emission associated with the AGN forms an ionization cone which is aligned with the nuclear radio jet. The AGN bolometric luminosity calculated from the AGN component of the \OIII\ emission is in close agreement with the bolometric luminosity calculated from the 2-10 keV X-ray emission.}
\item{The star formation component traces the \mbox{B-band} stellar continuum emission of the galaxy. The SFR peaks in the south-eastern region of the nuclear star forming ring seen at radio frequencies. The total SFR for the nuclear region of NGC~613, calculated from the star formation component of the \Ha\ emission, is in close agreement with the SFR calculated from the 12$\mu$m WISE emission over the same region.}
\item{The peak of the line emission associated with shock excitation is cospatial with the peak of the nuclear radio jet emission and with regions of strong $H_2$ and \mbox{[Fe II]} emission and high ionized gas velocity dispersion \mbox{($\sigma \sim$~300 km s$^{-1}$)}. The regions with the highest shock fractions ($f_{shock} \sim$~30-50 per cent) are likely to trace the outer boundary of the AGN ionization cone where outflowing material may be interacting with the surrounding ISM.}
\end{itemize}

The ability to separate emission associated with star formation, shock excitation and AGN activity will facilitate investigations into the ISM conditions and the interplay between different ionization mechanisms in the most complex and rapidly evolving astrophysical systems in the universe. The \Ha\ luminosity of the star formation component can be directly converted to the SFR and SFR surface density, the \Ha\ luminosity of the shock component can be converted to the shock luminosity and the \OIII\ luminosity of the AGN component can be converted to the AGN bolometric luminosity. The ability to calculate all of these quantities within individual spatially resolved regions of galaxies will facilitate detailed investigations of how the star formation efficiency is impacted by the compression of gas along shock fronts and by the AGN ionizing radiation field. Our results highlight the power of combining traditional emission line ratio diagnostic diagrams with integral field data.

\section{Acknowledgements}
B.G. gratefully acknowledges the support of the Australian Research Council as the recipient of a Future Fellowship (FT140101202). L.K. and M.D. acknowledge the support of the Australian Research Council (ARC) through Discovery project DP130103925. Support for AMM is provided by NASA through Hubble Fellowship grant \#HST-HF2-51377 awarded by the Space Telescope Science Institute, which is operated by the Association of Universities for Research in Astronomy, Inc., for NASA, under contract NAS5-26555. JKB acknowledges support from the Australian Research Council Centre of Excellence for All-sky Astrophysics (CAASTRO), through project number CE110001020. The Australia Telescope Compact Array is part of the Australia Telescope, which is funded by the Commonwealth of Australia for operation as a National Facility managed by CSIRO. The National Radio Astronomy Observatory is a facility of the National Science Foundation operated under cooperative agreement by Associated Universities, Inc.

\bibliography{mybib}

\appendix

\section{ATCA Observations, Data Reduction and Processing}
NGC~613 was one of 18 S7 sources observed in two bands centred at $5.5$ and $9.0\,$GHz in all four Stokes parameters ($I$, $Q$, $U$, $V$) using the ATCA Compact Array Broadband Backend \citep[CABB;][]{Wilson11}. CABB provides a $2\,$GHz bandwidth split into $1\,$MHz channels for each observing frequency. We were allocated 12 hours of ATCA green time on 16 January 2015 in the $6\,$km array under the project code C2987. We observed PKS 1934-638 at the beginning of the observations as the primary flux and bandpass calibrator. The observations were designed to maximize the $uv$ coverage for dynamic range imaging and determining the antenna complex gains and polarization leakage correction.

We calibrated the data using {\tt MIRIAD}\footnote{http://www.atnf.csiro.au/computing/software/miriad/} \citep{Sault95} version 1.5.  We used {\tt atlod} with options {\tt birdie} and {\tt rfiflag} to flag the known radio frequency interference (RFI) and restricted the bandpass to the known range of good frequencies using {\tt uvaver}.  Using the primary calibrator we calibrated the bandpass using {\tt mfcal} and applied the calibration to the secondary calibrator.  We then automatically flagged the data using the task {\tt pgflag} developed by \citet{Offringa10} and then manually flagged using {\tt uvflag}.  The 2\,GHz bandwidth of our observations requires {\tt nbins=16} (128\,MHz sections) in {\tt gpcal} to calibrate the antenna gains and phases and the instrumental polarisation over smaller frequency bands.  Transfer of the calibration solutions to the targets was completed using {\tt gpcopy}.  We then automatically flagged the target observations using {\tt pgflag} and manually flagged with {\tt uvflag}.

We carried out the final imaging and self-calibration using standard routines in the Astronomical Image Processing System (AIPS). For each frequency dataset, the split file used for the imaging was created after binning the 2049 initial channels into 256 channels. Both phase-only and phase + amplitude self-calibration were carried out. The 4.6 and 8.1 GHz images were convolved with circular beams of size 3 arcsec, and flux density values below 4 sigma were blanked out before creating the final spectral index image. The 4.6~GHz image has a peak surface brightness of 13.03 mJy beam$^{-1}$, an RMS error of 0.23 mJy beam$^{-1}$ and a total flux density of 39.94 mJy. The~8.1 GHz image has a peak surface brightness of 7.31 mJy beam$^{-1}$, an RMS error of 0.15 mJy beam$^{-1}$ and a total flux density of 24.04 mJy. We did not detect any linear polarized flux density in either Stokes $Q$ or $U$ down to $5\sigma_{\rm QU} = 250\,\mu$Jy beam$^{-1}$.

\end{document}